\documentclass{article}
\usepackage{graphicx}
\usepackage{amsmath}
\usepackage{psfig}
\usepackage{amsfonts}
\usepackage{amssymb}

\begin{document}
Dirac quantization condition with the superconducting state

\bigskip

Jer Yu Lin*

Department of physics, National Taiwan University, Taipei 10617,

Taiwan, Republic of China

Anthrop-Celestial Research Coordinate Institute, Tienti Church,

Nantou 55542, Taiwan, Republic of China

*Email address : r89222013@ms89.ntu.edu.tw\ \ 

\bigskip

\ \ \ The author argues that the Dirac quantization condition $u_{0}%
=\frac{\hbar c}{2e}$ might

imply the existence of an undiscovered electromagnetic structure which

governs the quantization of the electric charge and the quantization of the

magnetic flux in the superconducting state. An experimental set-up which

can provide a strong evidence by predicting the discrimination between

the magnetic flux generated by the positive and negative electric charge in

the superconducting state is also proposed.

\bigskip

\ \ \ \ Although Dirac had brilliantly elucidated the quantization condition between

the magnetic and the electric charge decades ago [1], the magnetic monopole

is still wanted, and ongoing efforts are invested in [2].Based on Dirac%
\'{}%
s

argument [1], the magnetic monopole $u_{0}$ was assigned to be the source of the

magnetic field obeying inverse-square law analogous to the role of the electric

charge q in the electric field. It is obvious to notice that the vector potential

needed to describe the magnetic field of $u_{0}$ can not do without a
singularity [3].

In order to better solve this problem, Wu and Yang proposed a pair of vector

potentials with the singularities lying along the -Z axis and the +Z axis

respectively [3]:

A$_{\phi}^{(1)}=\frac{u_{0}(1-\cos\theta)}{r\sin\theta}\ \ (0\leq\theta
<\pi-\varepsilon,\varepsilon\rightarrow0),$ \ \ \ \ \ \ \ \ \ \ \ \ \ \ \ \ \ \ \ \ \ \ \ \ \ \ \ \ \ \ \ \ \ \ \ \ \ \ \ \ \ \ \ \ (1)

A$_{\phi}^{(2)}=\frac{-u_{0}(1+\cos\theta)}{r\sin\theta}\ \ (\varepsilon
<\theta\leq\pi),$\ \ \ \ \ \ \ \ \ \ \ \ \ \ \ \ \ \ \ \ \ \ \ \ \ \ \ \ \ \ \ \ \ \ \ \ \ \ \ \ \ \ \ \ \ \ \ \ \ \ \ \ \ \ \ \ \ \ \ \ \ (2)

The region ($\varepsilon<\theta<\pi-\varepsilon$ ) where overlap of these two
vector potentials

happens signifies a gauge transformation leading to the Dirac quantization

condition [3]:

$u_{0}=\frac{\hbar c}{2e}$ (e = $\pm$1.602$\times10^{-19}$ coulomb, so in this
article I define +$u_{0}=\frac{\hbar c}{2e^{+}}$

, -$u_{0}=\frac{\hbar c}{2e^{-}}$),\ \ \ \ \ \ \ \ \ \ \ \ \ \ \ \ \ \ \ \ \ \ \ \ \ \ \ \ \ \ \ \ \ \ \ \ \ \ \ \ \ \ \ \ \ \ \ \ \ \ \ \ \ \ \ \ \ \ \ \ \ \ \ \ \ \ \ \ \ \ \ \ \ \ \ \ \ \ \ \ \ \ \ \ \ \ \ \ \ \ \ \ (3)

\ \ \ \ This well chosen gauge transformation inspired me to consider the situation

stated by Dirac [1] where the magnetic monopoles of opposite sign bound so

strongly that we had not observed any separated ones. Namely, the magnetic

fields exerted just mutually neutralize, making them become invisible. In terms

of the vector potential, we can write down two pairs of vector potentials for

this kind of situation:

1. A$_{\phi}^{(1)}$ = 0,\ \ \ \ \ \ \ \ \ \ \ \ \ \ \ \ \ \ \ \ \ \ \ \ \ \ \ \ \ \ \ \ \ \ \ \ \ \ \ \ \ \ \ \ \ \ \ \ \ \ \ \ \ \ \ \ \ \ \ \ \ \ \ \ \ \ \ \ \ \ \ \ \ \ \ \ \ \ \ \ \ \ \ \ \ \ \ \ \ \ \ \ \ \ \ (4)

\ \ \ A$_{\phi}^{(2)}=\frac{\pm2u_{0}}{r\sin\theta}$ $\ $($\varepsilon
<\theta<\pi-\varepsilon$ ),\ \ \ \ \ \ \ \ \ \ \ \ \ \ \ \ \ \ \ \ \ \ \ \ \ \ \ \ \ \ \ \ \ \ \ \ \ \ \ \ \ \ \ \ \ \ \ \ \ \ \ \ \ \ \ \ \ \ \ \ \ (5)

\bigskip B=$\nabla\times$A$_{\phi}^{(1)}=\nabla\times$A$_{\phi}^{(2)}=0$
$\ (\varepsilon<\theta<\pi-\varepsilon$ ), thus they are classically invisible.

$\oint$A$_{\phi}^{(2)}\cdot dl=\frac{hc}{e^{\pm}},$ \ \ \ \ \ \ \ \ \ \ \ \ \ \ \ \ \ \ \ \ \ \ \ \ \ \ \ \ \ \ \ \ \ \ \ \ \ \ \ \ \ \ \ \ \ \ \ \ \ \ \ \ \ \ \ \ \ \ \ \ \ \ \ \ \ \ \ \ \ \ \ \ \ \ \ \ \ \ \ \ \ \ \ \ \ \ (6)

which renders $\pm2\pi\ $to the associated Aharonov-Bohm phase of point charges

[4], making them quantum mechanically invisible as well. The gauge

transformation between them leads to the Dirac quantization condition.

2. A$_{\phi}^{(1)}$ = $\frac{+u_{0}}{r\sin\theta}$ $\ (\varepsilon<\theta
<\pi-\varepsilon$ ),\ \ \ \ \ \ \ \ \ \ \ \ \ \ \ \ \ \ \ \ \ \ \ \ \ \ \ \ \ \ \ \ \ \ \ \ \ \ \ \ \ \ \ \ \ \ \ \ \ \ \ \ \ \ \ \ \ \ \ \ \ (7)

\ \ \ A$_{\phi}^{(2)}=\frac{-u_{0}}{r\sin\theta}$ $\ $($\varepsilon<\theta
<\pi-\varepsilon$ ),\ \ \ \ \ \ \ \ \ \ \ \ \ \ \ \ \ \ \ \ \ \ \ \ \ \ \ \ \ \ \ \ \ \ \ \ \ \ \ \ \ \ \ \ \ \ \ \ \ \ \ \ \ \ \ \ \ \ \ \ \ (8)

which are constructed from the symmetrical consideration : A$_{\phi}^{(1)}$ is contributed

by + $u_{0}$, while A$_{\phi}^{(2)}$ is contributed by - $u_{0}$. B =
$\nabla\times A_{\phi}^{(1)}=\nabla\times A_{\phi}^{(2)}=0$ $\ $

$(\varepsilon<\theta<\pi-\varepsilon$ ), thus they are classically invisible.

$\oint A_{\phi}^{(1)}\cdot dl=\frac{hc}{2e^{+}}$ \ \ \ \ \ \ \ \ \ \ \ \ \ \ \ \ \ \ \ \ \ \ \ \ \ \ \ \ \ \ \ \ \ \ \ \ \ \ \ \ \ \ \ \ \ \ \ \ \ \ \ \ \ \ \ \ \ \ \ \ \ \ \ \ \ \ \ \ \ \ \ \ \ \ \ \ \ \ \ \ \ \ \ \ \ \ (9)

and

$\oint A_{\phi}^{(2)}\cdot dl=\frac{hc}{2e^{-}},\ $\ \ \ \ \ \ \ \ \ \ \ \ \ \ \ \ \ \ \ \ \ \ \ \ \ \ \ \ \ \ \ \ \ \ \ \ \ \ \ \ \ \ \ \ \ \ \ \ \ \ \ \ \ \ \ \ \ \ \ \ \ \ \ \ \ \ \ \ \ \ \ \ \ \ \ \ \ \ \ \ \ \ \ (10)

which render $\pm2\pi\ $to the associated Aharonov-Bohm phase of electron pairs

(q = 2e$^{-}$) in the superconducting state, making them quantum mechanically

invisible in the superconducting state. The gauge transformation between them

also leads to the Dirac quantization condition.

\ \ \ In other words, in an area where no magnetic monopole exists, these two

pairs of vector potentials still can reasonably exist except for the singularities.

Would they have any physical significance? So long as we have realized that

the Dirac quantization condition can be derived from the gauge transformation

represented by each of these two pairs of vector potentials, reversely,

if the Dirac quantization condition is held, we can get the Aharonov-Bohm

phase factor $f$ [3]:

$\pm$2$u_{0}f=1\rightarrow f=\frac{e^{\pm}}{\hbar c}$ $\ $(for the point
charges), \ \ \ \ \ \ \ \ \ \ \ \ \ \ \ \ \ \ \ \ \ \ \ \ \ \ \ \ \ \ \ \ \ \ \ \ \ \ \ \ \ (11)

$\pm u_{0}f=1\rightarrow f=\frac{2e^{\pm}}{\hbar c}$ $\ $(for the
superconducting state), \ \ \ \ \ \ \ \ \ \ \ \ \ \ \ \ \ \ \ \ \ \ \ \ \ \ \ \ \ (12)

\ \ \ \ Therefore, I assume that $\pm$2$u_{0}(\pm u_{0})$ just represents the
existence of one

hidden electromagnetic structure \ we have not discovered yet, it governs the

quantization of the electric charge (the quantization of the magnetic flux in the

superconducting state). The singularities in the vector potentials simply

represent the need of one unknown electromagnetic coordinate transformation

from the unknown electromagnetic structure represented by\ $\pm$2$u_{0}(\pm
u_{0})$ to the

one with quantized electric charges which we are familiar with.

\ \ \ \ Here I will not discuss the unknown electromagnetic coordinate

transformation, but I will discuss if the unknown electromagnetic structure

represented by $\pm$2$u_{0}(\pm u_{0})$ truly governs the quantization of the electric

charge (the quantization of the magnetic flux in the superconducting state),

could we have some experimental evidences?

1. The quantization of the electric charge : According to my assumption, the

electromagnetic energy goes through the unknown electromagnetic structure

represented by $\pm$2$u_{0}$ to the quantized electric charge $e^{\pm}$ =
$\pm$1.602$\times10^{-19}$

coulomb revealed by (11) and (6).

\ \ \ \ So the famous number 1/137 may be interpretated as follows:

$\pm$2$u_{0}$ = $\frac{\hbar c}{e^{\pm}}\rightarrow\pm$2$u_{0}$ /137 =
e$^{\pm}$ [1],\ \ \ \ \ \ \ \ \ \ \ \ \ \ \ \ \ \ \ \ \ \ \ \ \ \ \ \ \ \ \ \ \ \ \ \ \ \ \ \ \ \ \ \ \ \ \ \ \ \ \ \ \ \ \ \ \ (13)

which may represent the ratio for the electromagnetic energy quantized to the

electric charge e$^{\pm}$ to that still stays in the electromagnetic structure represented

by $\pm$2$u_{0}$, exactly as the coupling constant for the electromagnetic
energy to

an electron (positron), probably being a new interpretation of the famous

number 1/137.

2. The quantization of the magnetic flux in the superconducting state:

According to my assumption, the electromagnetic energy goes through the

unknown electromagnetic structure represented by $\pm u_{0}$ to the quantized

magnetic flux $\pm\Phi_{0}=\frac{hc}{2e^{\pm}}=\pm2.067\times10^{-15}weber$
revealed by (12) , (9)

and (10).

\ \ \ \ From the present knowledge of superconductivity, the magnetic flux in the

superconducting state is quantized to $\pm\Phi_{0}$, which had already been
verified in

experiments [5]. By conventional wisdom, the $\pm$ sign of the magnetic flux

merely indicates the incoming or outgoing of the magnetic flux. But in my

interpretation, since the $\pm$ sign of $\pm$2$u_{0}$ discriminates the
positive from the

negative sign of the electric charge, the $\pm$ sign of $\pm u_{0}$ should
discriminate the

positive from the negative sign of $\Phi_{0},$ namely +$\Phi_{0}$ generated by
the positive

electric charge from -$\Phi_{0}$ generated by the negative electric charge. In other

words, the magnetic flux generated by the positive (negative) electric charge

goes through the unknown electromagnetic structure represented by $+u_{0}$

($-u_{0}$) to the quantized magnetic flux $+\Phi_{0}$ $(-\Phi_{0})$. According
to eq.(12), the

Aharonov-Bohm phase factor for $+u_{0}$ ($-u_{0}$) is $\frac{2e^{+}}{\hbar c}$
($\frac{2e^{-}}{\hbar c}$), hence we have

achieved an important conclusion : the Aharonov-Bohm phase factor in the

superconducting state for the magnetic flux generated by the positive

(negative) electric charge is $\frac{2e^{+}}{\hbar c}$ ($\frac{2e^{-}}{\hbar
c}$), namely, for the associated

Aharonov-Bohm phase $\gamma$ of electron pairs in the superconducting state,

$\gamma=\gamma_{0}+\frac{2e^{-}}{\hbar c}\int A\cdot dl$\ \ (for the vector
potential generated by the negative

electric charge)\ ,\ \ \ \ \ \ \ \ \ \ \ \ \ \ \ \ \ \ \ \ \ \ \ \ \ \ \ \ \ \ \ \ \ \ \ \ \ \ \ \ \ \ \ \ \ \ \ \ \ \ \ \ \ \ \ \ \ \ \ \ \ \ \ \ \ \ \ \ \ \ \ \ \ \ \ \ \ \ \ \ \ \ \ (14)

$\gamma=\gamma_{0}+\frac{2e^{+}}{\hbar c}\int A\cdot dl$ (for the vector
potential generated by the positive

electric charge),\ \ \ \ \ \ \ \ \ \ \ \ \ \ \ \ \ \ \ \ \ \ \ \ \ \ \ \ \ \ \ \ \ \ \ \ \ \ \ \ \ \ \ \ \ \ \ \ \ \ \ \ \ \ \ \ \ \ \ \ \ \ \ \ \ \ \ \ \ \ \ \ \ \ \ \ \ \ \ \ \ \ \ \ (15)

this could be judged by the experiment.

\ \ \ \ The experiment could be done by utilizing the dc Josephson effect [6]:

I = I$_{0}\sin\gamma$ , \ \ \ \ \ \ \ \ \ \ \ \ \ \ \ \ \ \ \ \ \ \ \ \ \ \ \ \ \ \ \ \ \ \ \ \ \ \ \ \ \ \ \ \ \ \ \ \ \ \ \ \ \ \ \ \ \ \ \ \ \ \ \ \ \ \ \ \ \ \ \ \ \ \ \ \ \ \ \ \ \ \ \ \ \ \ \ \ \ (16)

by conventional wisdom, $\gamma=\gamma_{0}+\frac{2e^{-}}{\hbar c}\int_{1}%
^{2}A\cdot dl$ either for the vector potential

generated by the negative or positive electric charge, but in my interpretation,

the Aharonov-Bohm phase shift should be opposite in sign according to (14)

and (15). By intergrating eq.(16), we could get the Josephson junction diffraction

equation for a single Josephson junction [7]:

I$_{\max}=I_{c}\frac{\sin(\pi\Phi/\Phi_{0})}{\pi\Phi/\Phi_{0}},$\ \ \ \ \ \ \ \ \ \ \ \ \ \ \ \ \ \ \ \ \ \ \ \ \ \ \ \ \ \ \ \ \ \ \ \ \ \ \ \ \ \ \ \ \ \ \ \ \ \ \ \ \ \ \ \ \ \ \ \ \ \ \ \ \ \ \ \ \ \ \ \ \ \ \ \ \ \ (17)

which is plotted as figure1 [7].%
\begin{figure}
[ptb]
\begin{center}
\includegraphics[
height=2.0012in,
width=3.979in
]%
{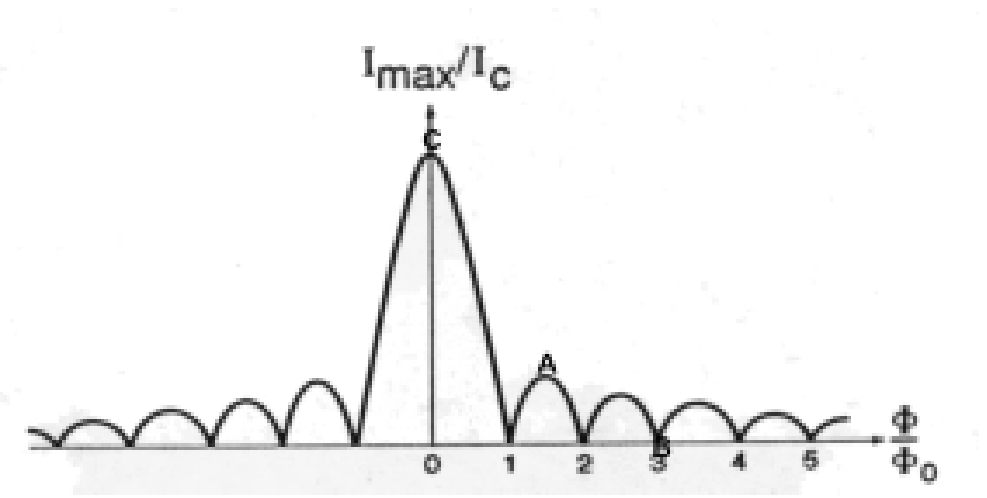}%
\end{center}
\end{figure}
\ \ \ \ \ \ \ \ \ \ \ \ \ \ \ \ \ \ \ \ \ \ \ \ \ \ \ \ \ \ \ \ \ \ 

\bigskip\ \ \ \ \ \ \ \ \ \ \ \ \ \ \ \ \ \ \ \ \ \ \ \ \ \ \ \ \ \ \ \ \ \ \ \ \ figure1

\ Therefore, if we set up a single Josephson junction under the static magnetic

flux generated by the positive charge (ex. a zinc coil whose charge carrier

is positive) to arrive at, say point A in figure1, then we add completely the

same static magnetic flux generated by the negative charge (ex. a copper

coil whose charge carrier is negative) onto it, we will arrive at point B by

conventional wisdom, but according to my interpretation, we should arrive

at point C.

\ \ \ If the above experiment could be verified, we could attain a strong evidence

that there truly exists an electromagnetic structure represented by $\pm
$2$u_{0}(\pm u_{0})$

which governs the quantization of the electric charge (the quantization of the

magnetic flux in the superconducting state). We might understand more about

the number 1/137 and the transition between the normal state and the

superconducting state ($\pm$2$u_{0}\longleftrightarrow\pm u_{0}$) if we go
deep into the study of this structure.

\bigskip\ 

*Email address : r89222013@ms89.ntu.edu.tw\ \ 

[1] P. A. M. Dirac, Proc. R. Soc. London A133, 60 (1931).

[2] J. D. Jackson, Classical Electrodynamics, John Wiley and Sons, Inc.,

\ \ \ \ \ New York, 1998, P.273.

[3] T. T. Wu and C. N. Yang, Phys. Rev. D 12, 3845 (1975).

\ \ \ \ \ J. J. Sakurai, Modern Quantum Mechanics, Addison-Wesley

\ \ \ \ \ Publishing Company, Inc., 1994, P.140-143.

[4] Y. Aharonov and D. Bohm, Phys. Rev. 115, 485 (1959).

[5] B. S. Deaver, Jr., and W. M. Fairbank, Phys. Rev. Lett 7, 43 (1961).

\ \ \ \ \ R. Doll and M. Nabauer, Phys. Rev. Lett 7, 51 (1961).

[6] B. D. Josephson, Physics Letters 1, 251 (1962).

[7] C. P. Poole, Jr., H. A. Farach, R. J. Creswick, Superconductivity,

\ \ \ \ \ Academic Press, Inc., 1995, P.441.
\end{document}